\def\kms{\relax \ifmmode {\,\rm km\,s}^{-1}\else \,km\,s$^{-1}$\fi}   
  
\def\farcs{\hbox{$.\!\!^{\prime\prime}$}}

\def\secd#1.#2{ #1\farcs#2 }               

\def\mincir{\ \raise-2.truept\hbox{\rlap{\hbox{$\sim$}}\raise5.truept  
    \hbox{$<$}\ }}  
\def\magcir{\ \raise-2.truept\hbox{\rlap{\hbox{$\sim$}}\raise5.truept  
    \hbox{$>$}\ }}

\def\nii{[N~{\sc ii}]}  
\def\oiii{[O~{\sc iii}]}  

\def\oii{[O~{\sc ii}]}  
\def\ha{H$\alpha$} 
\def\hb{H$\beta$}

\def\ariii{[Ar~{\sc iii]}}  
  
\def\hii{H~{\sc ii}}
\def\hi{H~{\sc i}}  
\def\heii{He~{\sc ii}}

\def\hei{He~{\sc i}} 
\def\sii{[S~{\sc ii}]}

\documentclass[referee]{aa} 
\usepackage[dvips]{graphics}    
\begin{document}  
\title{Chemical Abundances of Planetary Nebulae in M33\thanks{Based on 
observations obtained at the 4.2m~WHT telescope  
operated on the island of La Palma by the Isaac Newton Group in the Spanish  
Observatorio del Roque de Los Muchachos of the Instituto de Astrofisica de  
Canarias.}}  
 
\author{ 
L. Magrini     \inst{1}, 
M.   Perinotto \inst{1},  
A.   Mampaso   \inst{2},
R.L.M. Corradi \inst{3} 
 }  
\offprints{L. Magrini\\   
e-mail: laura@arcetri.astro.it}  
\institute{ 
Dipartimento di Astronomia e Scienza dello Spazio, Universit\'a di 
Firenze, L.go E. Fermi 2, 50125 Firenze, Italy 
\and  
Instituto de Astrof\'{\i}sica de Canarias, c. V\'{\i}a L\'actea s/n, 
38200, La Laguna, Tenerife, Canarias, Spain 
\and   
Isaac Newton Group of Telescopes, Apartado de Correos 321, 38700 Santa   
Cruz de La Palma, Canarias, Spain
} 
 
\date{Received date /Accepted date }  
 
\abstract{
Using spectroscopic data presented in Magrini et al. (\cite{m03}), we
have analyzed with the photoionization code CLOUDY 94.00 (Ferland et
al. \cite{ferland})  11 Planetary Nebulae belonging to the spiral
galaxy M~33.  Central star temperatures and nebular parameters have been
determined. In particular the chemical abundances of He/H, O/H, N/H,
Ar/H, and S/H have been measured and compared with values obtained via
the Ionization Correction Factors (ICFs) method, when available.
Chemical abundance relationships have been investigated; in
particular, a correlation between N/H and N/O similar to the Galactic
one (Henry \cite{henry}), and a feeble anti-correlation between O/H and N/O 
have been found.  
 A gradient in O/H across the disc of M~33 is  
indicatively consistent with the one found from \hii\ regions 
in this galaxy (V\'{\i}lchez et al \cite{vilchez88}). 
Further studies in the more external parts of M~33 are however needed to ascertain this point.
The present result shows that oxygen and helium abundances (with lower
accuracy also nitrogen, argon and sulphur) can be actually estimated
from the brightest PNe of a galaxy, even if the electron temperature
cannot be measured.  We also found that the oxygen abundance is quite
independent of the absolute magnitude of the PN and consequently the
brightest PNe are representative of the whole PN population.  This
represents an important tool to measure the metallicity of galaxies at
the time of the formation of PNe progenitors.
\keywords{Planetary nebulae:individual: M33 -- Galaxies:individual: M33 --
Galaxies: Abundances} }
\authorrunning{Magrini et al.}   
\titlerunning{Chemical Abundances of Planetary Nebulae in M~33}    
\maketitle   
  
\section{Introduction}

Chemical abundances in extragalactic Planetary Nebulae (PNe) have been
derived so far only in a small number of galaxies.  Inside the Local
Group they are: the irregular galaxies LMC, SMC (cf. Leisy et al.
\cite{leisy}; Jacoby \& De Marco \cite{jacobydemarco}), and NGC~6822 
(Richer \& McCall \cite{richer}), the dwarf spheroidal and spheroidal
galaxies Fornax (Maran et al. \cite{maran}), Sagittarius (cf. Walsh et
al. \cite{walsh}), NGC~185, and NGC~205 (Richer
\& McCall \cite{richer}), the elliptical galaxy M~32 (Stasinska et al. 
\cite{stasinska}) and finally the two spiral galaxies M~31 and M~33 that
have been studied by Jacoby \& Ciardullo (\cite{jc99}, hereafter JC99)
and Magrini et al. (\cite{m03}, hereafter M03), respectively.  There
is a substantial lack of information about chemical abundances of PNe in
dwarf irregular galaxies of the Local Group
(e.g. Sextans~B. Sextans~A, Leo~A, IC~10), where no PN has been
investigated so far. 
Outside the LG, only PNe in the giant
elliptical galaxy NGC~5128 (Centaurus~A) at the distance of 3.5 Mpc
were analyzed by Walsh et al. (\cite{walsh99}).

In most of the previous cases, with the exception of NGC~5128, the
abundances have been obtained by direct determination of the T$_e$,
derived from the \oiii\ $\lambda\lambda$ 4363/5007 \AA\ or, in the worst
cases when the $\lambda$ 4363 \AA\ is not measured, using its upper limit, thus
obtaining lower limits for the \oiii\ ionic abundance and then for
the total O abundance, which was calculated from the ICFs (Ionization
Correction Factors) procedure (cf. e.g. Kingsburgh \& Barlow
\cite{kb94}, hereafter KB94).

In M~31, in addition to the ICFs method, JC99 have modeled the
observed PNe using the photoionization code CLOUDY 90 (Ferland et
al. \cite{ferland}) with simplified assumptions: blackbody central
stars and spherical nebulae with constant density.  They have derived
abundances and central star parameters for 15 PNe of M~31. In eight
cases they could compare the results from the nebular models built
with CLOUDY with abundances based on the ICFs method. They found good
agreement between abundances derived with these two different
methods. From this comparison, CLOUDY resulted to be a useful tool to
measure chemical abundances, particularly when direct electron
temperature measurements are not available.  

M03 studied  48 emission line objects in M~33 spectroscopically, and recognized,
via diagnostic diagrams, 26 of them to be {\sl bona fide} PNe.  They
presented the observed fluxes for 42 of the 48 objects, deriving
chemical abundances with the ICFs method for the three brightest PNe.

In the present paper, we re-analyze with CLOUDY 94.00 (Ferland et
al. \cite{ferland}) 11 of the 26 PNe presented by M03 using the
available spectroscopic data.  The three PNe already studied with ICFs
by M03 are also included, in an effort to clarify their observed low
N/O.

In Section 2 we summarize the observations and data reduction.  In
Section 3 the derivation of nebular and stellar properties is
illustrated.  In Section 4 we discuss the derived quantities 
and correlations among them. Summary and Conclusions are in
Section 5.

\section{Observations and data reduction}

The observations were taken at the 4.2 WHT telescope (La Palma, Spain)
in October, 2001 with AF2/WYFFOS, with a setting producing a
dispersion of 3.0 \AA\ per pixel in the spectral range  4300 \AA\ to
7380 \AA.  A detailed description of spectroscopic data and of the
reduction procedure are in M03.
\begin{figure} [h!]
\resizebox{\hsize}{!}{\includegraphics{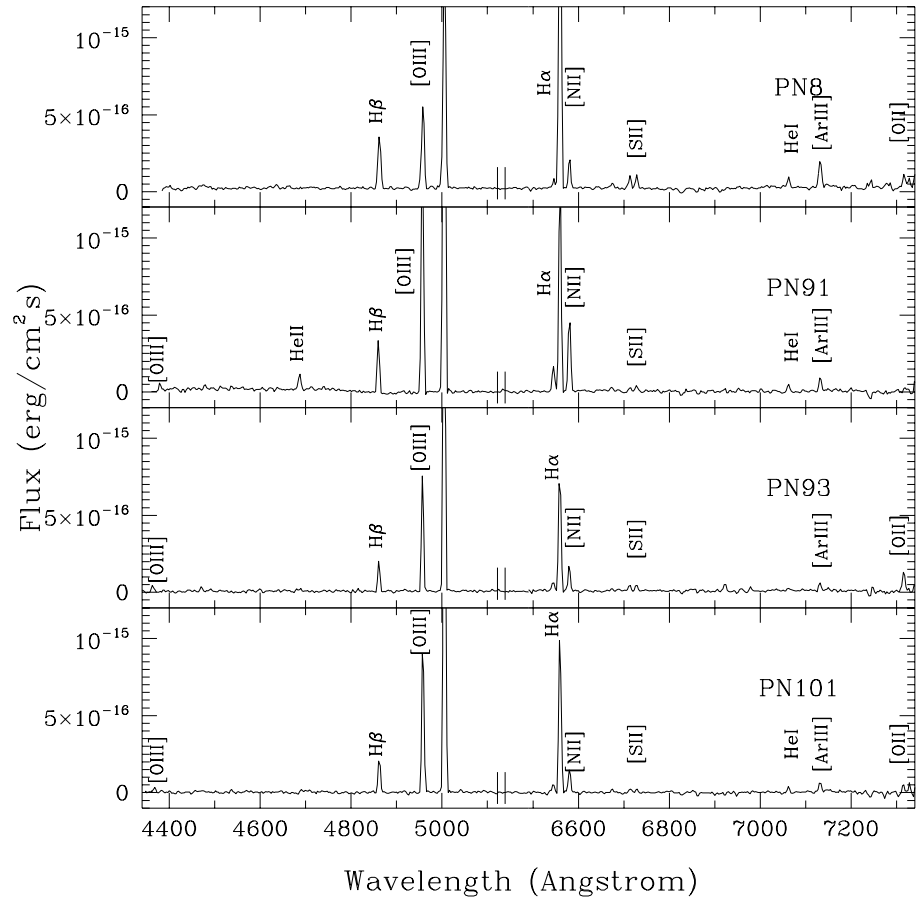}} 
\caption{ Characteristic spectra of PNe in M~33 obtained 
by M03 with an exposure time of 8.7~hrs with the WHT.  } 
\label{fig0}
\end{figure}  

\section{The modelling procedure}

Using a photoionization model such as CLOUDY, it is possible to overcome
the lack of direct information on the electron temperature in a PN,
due to the faintness of the relevant forbidden lines: the  \oiii\ $\lambda$4363
 and the \nii\ $\lambda$5755 emission lines, which are hard to detect in
extragalactic PNe.  On the other hand, the use of CLOUDY requires 
 as input the energy distribution of the ionizing radiation of the 
central star, its luminosity and the nebular geometry.
The distance to the object must then be known.
 In our case the distance is available adopting for all PNe in M~33 the distance
of the galaxy, equal to 840$\pm$90~kpc (Magrini et al. \cite{m00}, hereafter M00).
As with the energy distribution of central stars,  the usual
simplification is to assume them to behave as  blackbodies (BB), while 
 for a better accuracy stellar atmosphere models are needed (cf. Morisset \cite{morisset}).
 In this respect, an illustrative test was carried out by Howard et al. (\cite{howard}). 
They tested the effects on derived abundances and central 
star temperatures of a variety of model atmospheres and blackbodies  
in a set of Galactic halo PNe.
They found that nebular line strengths are relatively 
insensitive to atmospheric model details and that blackbody spectra are by and large  suitable 
to represent the central star continua.
On the other hand Armsdorfer et al. (\cite{armsdorfer}) checked in the particular case 
of the spatially resolved round Galactic PN NGC~2438, the 
effects of modelling a PN central star with a blackbody  instead than  
with a  model atmosphere. They found that the BB assumption  leads to  
underestimating   the \oiii\  $\lambda$5007 line and 
to overestimating  the \heii\  $\lambda$4686 line strength.
 Therefore the assumption of  BB central stars might introduce  
uncertainties in the determination of chemical abundances. 
We will also check in the next sections in a specific case the 
BB assumption vs. the use of a model atmosphere. 
We will see that differences in resulting abundances are 
quite small.

The aim of our procedure is to match the intensity of the observed
emission lines with that predicted by CLOUDY. The spirit of our
procedure is similar to the one adopted by JC99 to determine the
chemical abundances of PNe in M~31. In the following we briefly
describe how our analysis works.

CLOUDY needs  the following input parameters: the
central star energy distribution and luminosity, the nebular geometry,
the electron density, and the chemical abundances.  Because of the  lack of information 
in our objects on the morphology  and on the 
density distribution, we set the geometry to be spherical and  the density to a constant value, 
 although we will explore the effect of a $r^{-2}$ density distribution in some PNe.  
 The central star energy distribution was set to that of a BB.
For the three PNe (PN91, 93, 101) with the best S/N spectra and with  chemical abundances 
derived also with the ICFs method (M03), we will compare nebular abundances from 
the blackbody central star with those from the model atmospheres by Rauch (\cite{rauch}) (see Table\ref{tab2}).

The first run is done setting the input parameters as follows.  The
BB temperature is derived using the Ambartsumian's method
(Ambartsumian \cite{ambar}), i.e. using the
\heii(4686)/\hi(4861) line flux ratio, or the 
(\oiii(4959)+\oiii(5007))/\heii(4686)) lines ratio method (Gurzadyan \cite{gurzadyan})
for  medium and high excitation objects, respectively.  The  medium and high
excitation is recognized from the observed line ratio
\heii(4686)/\hi(4861). Following Gurzadyan (\cite{gur97})  we call high
excitation PNe  those with  \heii(4686)/\hi(4861) $>$ 0.15. \footnote{Gurzadyan (\cite{gurzadyan})
indeed distinguishes between low and high excitation objects. 
We prefer to use the term 'medium' in place of low because 
when the \heii\ $\lambda$4686 line is measured, the 
excitation cannot be really regarded as low. 
We think this specification is more appropriate to PNe where 
the line is not detectable.}

The central star luminosity is set to reproduce the observed absolute
flux of the \oiii\ $\lambda$5007  nebular line,  measured by
Magrini et al.  (\cite{m00}) via aperture photometry and 
reddening corrected  with E(B-V)=0.07 (van den Bergh (\cite{bergh})).
We made an effort to correct each $\lambda$ 5007 \AA\ \oiii\ flux with 
the individual reddening coming from the Balmer decrement. We saw 
that in some cases the corrected luminosity was substantially above 
the cutoff of the \oiii\ luminosity function. 
Thus we conclude that the de-reddening values from the Balmer decrement 
suffer from calibration issues. 
In these cases, while the line fluxes can be considered 
to be correct,  c$\beta$ is not appropriate  to correct the 
photometric fluxes. We have then preferred to correct the photometric fluxes
using the average extinction towards M~33. 
  
 We varied the external radius from 10 $^{16}$ cm (0.0032 pc) 
to 10 $^{18}$ cm (0.32 pc). This range  corresponds to 
a reasonable interval of sizes of Galactic PNe.
The electron density is evaluated from the \sii\ 
$\lambda\lambda$6717/6731 line ratio.  We have assumed the
average non-Type I galactic PNe abundances from KB94 as initial
chemical abundances  and dust grains, with composition and size considered to be 
typical of Galactic PNe dust (cf. Ferland \cite{ferland}).  
The addition of dust in the model does not vary substantially 
the derived chemical abundances. 

Further iterations make the following
adjustments.  First the BB T$_{\star}$ is altered to match the
\hei/\heii\ line ratio. 
 When the \heii\ $\lambda$4686 line was too weak to be measured, we determined  its 
upper limit so as to derive an upper limit to the central star temperature. These cases correspond 
to T$_{\star}$ marked with b in Table~\ref{tab1}, while directly measured stellar 
temperatures are marked with a. 
In the few cases, where  the \oii\ $\lambda$7325  doublet was measured, 
the nebular radius
was  varied to match the  observed ratio of the nebular lines
\oiii\ $\lambda$5007  and \oii\ $\lambda$ 7325 
(cases marked with c in Table~1) (cf. Che \& K$\ddot{o}$ppen \cite{che}) 
 In the other cases, models with different nebular radii 
were built. For these PNe, we found the smallest and the largest radius for which 
the model converged. Final quantities are the averages of quantities from models
spanning over these nebular radius intervals (cases marked with d in Table~1).

Subsequently chemical abundances are varied in
order to match the observed and predicted intensities within 5\% for
the  lines brighter than \hb\ and 20\% for the other lines.  
The  T$_{e}$ derived from the model is compared with the observed one, when available, 
i.e. in the three PNe where the \oiii\ 4363 \AA\ emission-line was measurable.
In these  PNe, the abundances of C and O (the most important
coolants) were slightly modified  to match the
observed T$_{e}$. The optical depth of the nebula is derived from the 
hydrogen ionization structure predicted by the CLOUDY model. 
The PNe are optically thick, while a few  with lower density (PN28, PN96, PN125)
 resulted to be  optically thin.

Generally, we need from 10 to 30 iterations to
reach  convergence, i.e. to reproduce within 5-20\% the observed line spectrum.
 Well determined quantities are  the luminosity 
of the central star, because distance and  absolute \oiii\ 5007 \AA\ flux 
are both well known, and the temperature of the central star (when both \hei\ and \heii\ 
lines are measured). 
The  determination of the nebular radius is
instead weaker, because the measurement of $\lambda$ 7325 \AA\ \oii\
doublet is affected by large observational errors. 
 In two PNe where  the \oii\  $\lambda$7325 doublet was measured 
(PN  91, 93) we introduced the dependence on r$^{-2}$ in the density to find 
better agreement between observed and predicted \oiii/\oii\ ratio.   
In fact  the  ratio \oiii/\oii\  is not only sensitive to the size of the nebula, but it depends 
also on several other parameters, mainly on the density distribution  (cf. Che \& K$\ddot{o}$ppen \cite{che}). 
 When the central star temperature is derived from 
the upper limit of the \heii\ $\lambda$4686 line, only upper limit to 
the helium and other metal abundances can be derived. 

The derived central
star temperatures and nebular quantities are shown in Table~\ref{tab1}. 
We recall that the  T$_{\star}$  derived from the upper limit of the \heii\ $\lambda$4686  are marked with b, 
whereas the directly measured T$_{\star}$ are marked with a. 
Derived nebular radii are marked with c, and  upper limit nebular radii with d. 

In Table~\ref{tab2} chemical
abundances for three PNe computed using ICFs (M03) are compared with
current CLOUDY results. 
Note that for Ar/H the ionization correction factors are larger that 
one because only the \ariii\ $\lambda$7135 line is detected.

\subsection{Tests of the procedure}

\begin{figure} [h!]
\resizebox{\hsize}{!}{\includegraphics{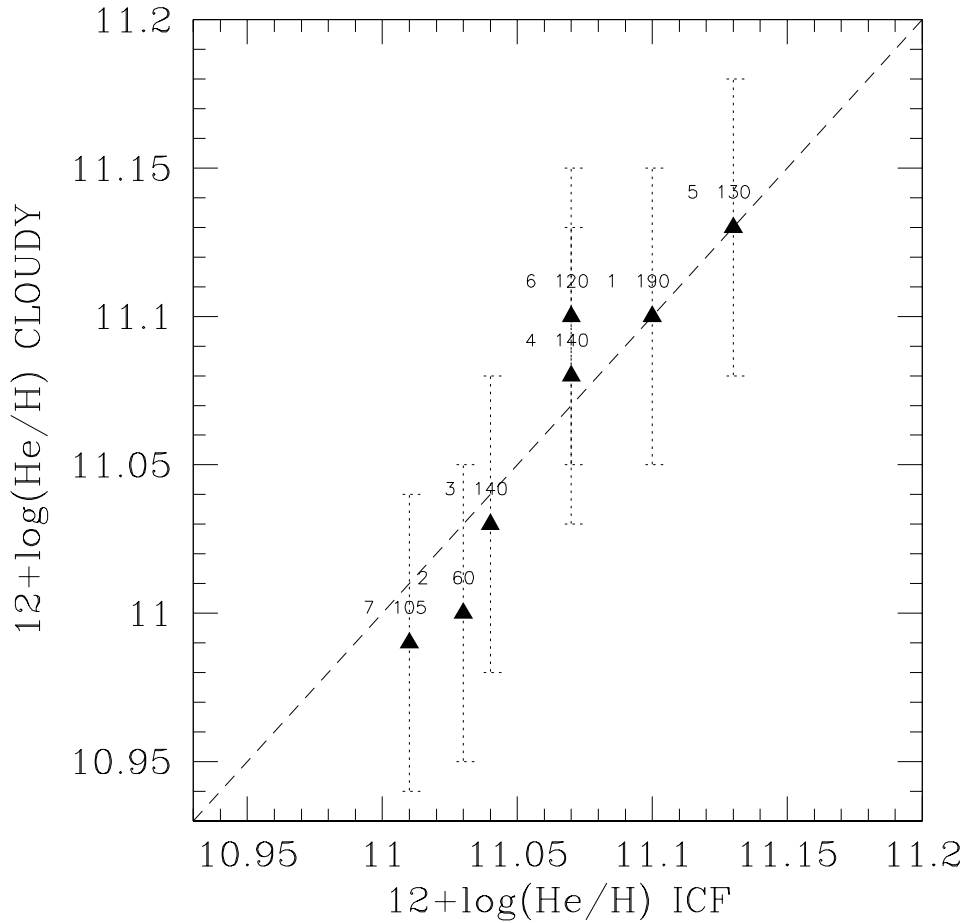}} 
\caption{Comparison of He/H abundance of a sample of Galactic PNe derived with ICFs method by KB94
with He/H computed using CLOUDY.
Triangles represent He/H when T$_{\star}$ has been  derived in the present work.
The id numbers refer to the Galactic PNe: 1--M1-8, 2--M3-1, 3--M3-5, 4--NGC2867, 5--NGC3195, 6--NGC7009, 7--NGC5852. 
The derived T$_{\star}$/1000 is reported near to 
the corresponding measured He/H. } 
\label{fig1}
\end{figure}

\begin{figure}[h!]
\resizebox{\hsize}{!}{\includegraphics{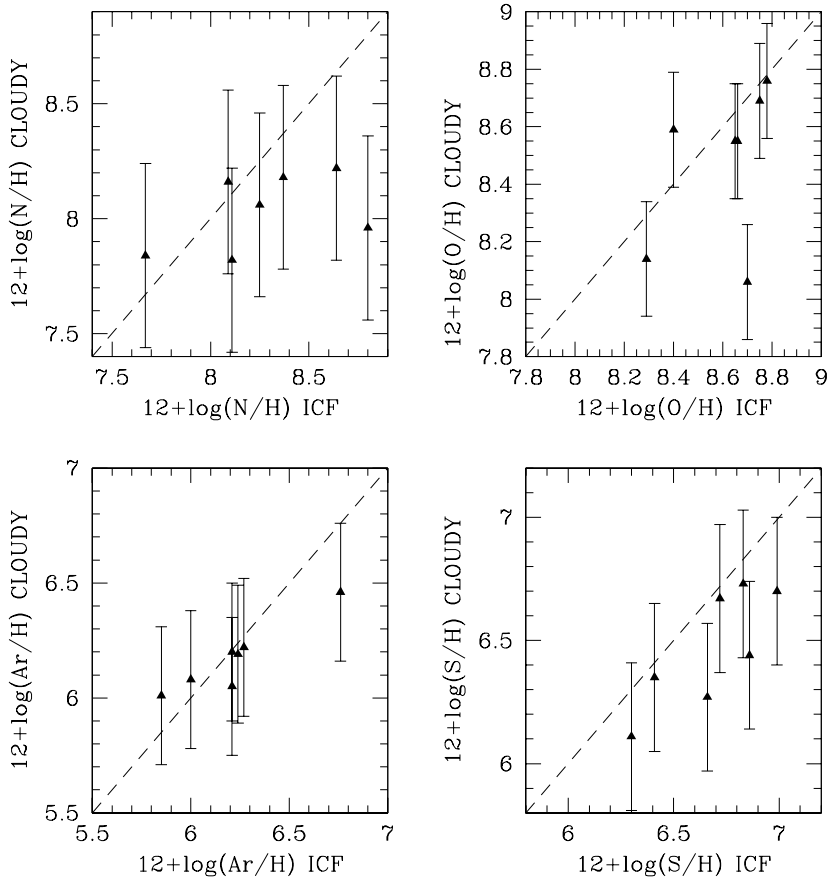}} 
\caption{Comparison of O/H, N/H, S/H, and Ar/H abundances of seven Galactic 
PNe derived with the ICFs method by KB94 
with the same quantities computed using CLOUDY.} 
\label{fig2}
\end{figure}

We have tested the accuracy of the method by applying it to seven bright
Galactic PNe (M1-8, M3-1, M3-5, NGC2867, NGC3195, NGC7009, NGC5852), 
using only the emission lines present in our extragalactic
PN spectra, and comparing our results with the chemical abundances
obtained with the ICFs method by KB94.  PNe with central star 
BB temperatures spanning in a large range (from approximately 60,000 to 190,000
K) have been analyzed. 
 For these tests we have used the basic approximations of our model: 
BB central star and constant density. 
The results are shown in Figures
\ref{fig1} and \ref{fig2}.  We found a good agreement in the helium
abundance, with a r.m.s. difference of 0.015 dex (triangles in Fig.\ref{fig1}). 
The typical r.m.s. differences of the two determinations, with CLOUDY and with ICFs, of
O/H is 0.15 dex, of Ar/H is approximately 0.1 dex  and of N/H is $\sim$ 0.3 dex.  In some
cases these quantities are underestimated by our method whereas
sometimes are overestimated.  On the other hand S/H and N/H appear to be
generally underestimated by CLOUDY, with a r.m.s. difference of 0.2
dex. 
A possible explanation for the systematic differences in  S/H  and N/H 
determination might be related to the fact that for these two elements  only 
low ionization lines are observed. The considered PNe are medium and high excitation 
objects where these ions represent only a minor fraction of the total elemental 
abundance. Moreover the ionization potential of \sii\ is lower than 
that of \nii, which might explain the larger discrepancy in S/H.



\section{Chemical Abundances}

We have applied the procedure above described to 11 PNe of M~33.  In
 four  cases we could determine the temperature of the central star
within 5,000 K using \hei\ and \heii\ lines. In the remaining cases we
 estimated an upper limit to the central star temperature using 
upper limits to the \heii\ $\lambda$4686  flux.
 Considering the errors in the observed fluxes, the uncertainties in the model representation, and 
the quality in the fit  obtained,  we estimate typical errors are:
in He/H of 0.05 dex (0.10 dex when T$_{\star}$ is an upper limit), of 0.2 (0.3)
dex in O/H, and of 0.3 (0.4) dex in  Ar/H and S/H, and of 0.4 (0.5) dex in N/H. 
We note that S/H is generally under-estimated by the CLOUDY method in our test with Galactic PNe, and also N/H appears 
low compared to ICFs method for the highest values of N/H.

Chemical abundances, central star parameters, nebular radius and
electron density are shown in Table \ref{tab1}.
Chemical abundances previously obtained for the three PNe with the ICFs
method (M03) are shown in Table \ref{tab2} together with the new
determinations, both with BB and model atmosphere (Rauch \cite{rauch}).  
 The four  sets of determinations are consistent within the errors. 
Since the difference of the chemical abundances derived with CLOUDY 
using the two  stellar ionizing sources are rather small, 
we used for the remaining PNe a BB central star.
The use of a density dependent on r$^{-2}$ with a BB central star produces better agreement 
between predicted and observed \oiii/\oii\ ratio. With this model we 
obtain a higher N/H and a lower O/H, and consequently a higher N/O than with ICFs method.  


S/H has been re-derived because of a numerical mistake in the previous work (M03).  
The discrepancy
of S/H computed with ICFs and with CLOUDY (both BB and model atmosphere) 
is $\sim$-0.2 dex,confirming  the result mentioned  above that  CLOUDY tends to 
slightly underestimate  S/H.

The comparison in the brightest PNe of M~33 of abundances derived with the  ICFs and
those from photoionization models, confirms the accuracy of the CLOUDY procedure
in measuring helium, oxygen, nitrogen and argon abundances.


\begin{table*}

\caption{Central star temperatures and nebular parameters. Chemical abundances are expressed in 12+$\log$(X/H).
The line fluxes $F_{\rm H\alpha}$ and $F_{\rm [OIII]}$ are observed (M00). The adopted reddening correction is E(B-V)=0.07 
(van den Bergh (\cite{bergh})).}  
\label{tab1}
\begin{center}
{\scriptsize
\begin{tabular}{r r r r l l l l r r r r r r }    
\hline    
Id  & d & $F_{\rm H\alpha}$ & $F_{\rm [OIII]}$ & T$_{\star}$ & log(L$_{\star}$/L$_{\odot}$)     & log(r)  & log(N$_e$) & He/H & N/H & O/H & S/H & Ar/H & N/O \\
M00 & arcmin &\multicolumn{2}{c}{\footnotesize{10$^{-15}$erg/cm$^2$s}}&K/1000& 	&cm & cm$^{-3}$ &    &   &   &  &  &    \\
\hline  
8   &22 & 11.4 & 10.1 & 74~b  & 3.8 &16$<$r$<$18~d    & 3.1 &  11.05 &7.07 &7.94 &6.38 &6.11& 	-0.87	\\
18  &13 & 3.0  & 14.2 & 100~b & 4.1 &16$<$r$<$17.9~d  & 2.5 &  11.09 &7.48 &8.20 &6.83 &6.51&       -0.72\\
28  &12 & 2.9  & 13.9 & 142~a & 3.7 &16$<$r$<$17.7~d  & 2.1 &  10.96 &6.98 &8.69 &6.47 &6.3&       -1.71\\
60  &13 & 1.4  & 3.4  & 160~a  & 3.1 & 17~c           & 2.9 & 11.04 &7.67 &8.69 &6.57 &6.31&       -1.02\\
65  &8  & 10.1 & 5.6  & 128~b  & 3.2 & 16$<$r$<$17.7~d& 3.2 & 11.16 &7.61 &8.67 &6.84 &6.63&       -1.06\\
75  &11 & 4.4  & 12.5 & ~96~b  & 3.5 &16$<$r$<$17~d    & 3.9 & 11.14 &8.03 &8.49 &6.81  &6.41&       -0.46\\
91  &18 & 4.9  & 16.3 & 110~a  & 3.6 &17.0~c	      & 3.6 & 11.31 &8.07 &8.48 &6.60 &6.18&       -0.41\\
93  &16 & 3.1  & 7.4  & ~96~a  & 3.3 &17.2~c	      & 3.7 & 10.93 &7.75 &8.47 &6.66 &6.09&       -0.72\\
96  &9  & 3.1  & 11.9 & 105~b  & 3.5 &16$<$r$<$17.9~d & 1.9 & 11.06 &7.43 &8.76 &6.44 &6.33&       -1.33\\
101 &14 & 4.1  & 11.4 & ~97~b  & 3.4 & 17~c	      & 2.5 & 11.07 &7.33 &8.72 &6.18 &6.27&       -1.56\\
125 &9  & 7.5  & 8.2  & ~86~b  & 3.6 & 16$<$r$<$17.9~d& 2.2 & 10.93 &7.53 &8.31 &6.66 &6.14&       -0.78\\
\hline
\multicolumn{4}{l}{M~33 PNe all (1)}   	                && &  & & 11.08 &7.65  & 8.54 &6.62  &6.33 &-0.89\\
\multicolumn{4}{l}{M~33 PNe with T$_{\star}$ measured (1)}   &   & &  & & 11.09 &7.78  &8.61  &6.60  &6.28 &-0.83\\
\multicolumn{4}{l}{M~33 \hii\ regions (2)} &  &             &       &  & 10.92 &7.34  &8.55  &6.95  &-     & -1.21  \\
\hline
\multicolumn{4}{l}{Galactic PNe non-Type I (3)}  &               &  & & & 11.05 &8.14  & 8.69  &6.91  &6.38&-0.55\\
\multicolumn{4}{l}{Galactic PNe Type I (3)}  &               &      &&  & 11.12 &8.72  & 8.65  &6.91  &6.42&+0.07\\
\multicolumn{4}{l}{Galactic \hii\ regions (4)}         &    & &       &  & 11.00 &7.57  & 8.70  &7.06  &6.42&-1.13\\
\hline
\multicolumn{4}{l}{LMC PNe non-Type I (5)}                 & &     & &  & 10.96  &7.46   &8.35   &6.81   &5.95 &-0.90\\
\multicolumn{4}{l}{LMC PNe Type I (5)}                  &   &  &       & & 10.95  &8.28   &8.24   &7.11   &6.12 &+0.02\\
\multicolumn{4}{l}{LMC \hii\ regions (6)}    &    &      &   &         & 10.97  &6.97   &8.38   &6.67   &6.92 &-1.41\\

\hline
\multicolumn{5}{l}{(1) Present paper (average, by number, not by $\log$);}                    &      &      &       &  &  & &  &   \\
\multicolumn{4}{l}{ (2) V\'{\i}lchez et al. (\cite{vilchez88});}           &         &            &    &   &  &  & &  &   \\
\multicolumn{4}{l}{ (3) KB94; }        &      &         &      &            &  &  && &     \\
\multicolumn{4}{l}{ (4) Dufour (\cite{dufour84}) }     &      &               &      &   &      &  & &  &   \\
\multicolumn{4}{l}{ (5) Leisy \& Dennefeld (\cite{leisy04}) }     &              &      &  &          &  &  & &  &   \\
\multicolumn{4}{l}{ (6) Dennefeld (\cite{dennefeld}) }     &      &               &      &  &       &  & &  &   \\

\end{tabular}
 }
\end{center}
\end{table*}

Stellar evolution theory predicts that He and N are enhanced in
PNe by nucleosynthesis process, whereas elements heavier than N, like
O, Ar, S do not vary significantly from the moment of the formation of
their central stars (cf. Iben \& Renzini \cite{iben}).  In M~33, a
comparison of our determinations of the chemical abundances of PNe 
(particularly for PNe whose T$_{\star}$ could be measured, see Table \ref{tab1})
with those of \hii\ regions (V\'{\i}lchez et al \cite{vilchez88}), especially for the
best determined elements He/H and O/H (Table \ref{tab1}), confirms the
reliability of the determinations.  
In fact, as expected, He/H (and also N/H, but remember the larger errors that 
affect this determination) is
enhanced compared to \hii\ regions while O/H is almost unchanged.
This indicates that He and N have undergone nucleosynthesis enrichment in PNe progenitors,
as theory predicts.  Consequently N/O appears to be increased in the M~33 PNe
compared with \hii\ regions.

\begin{figure}
\resizebox{\hsize}{!}{\includegraphics{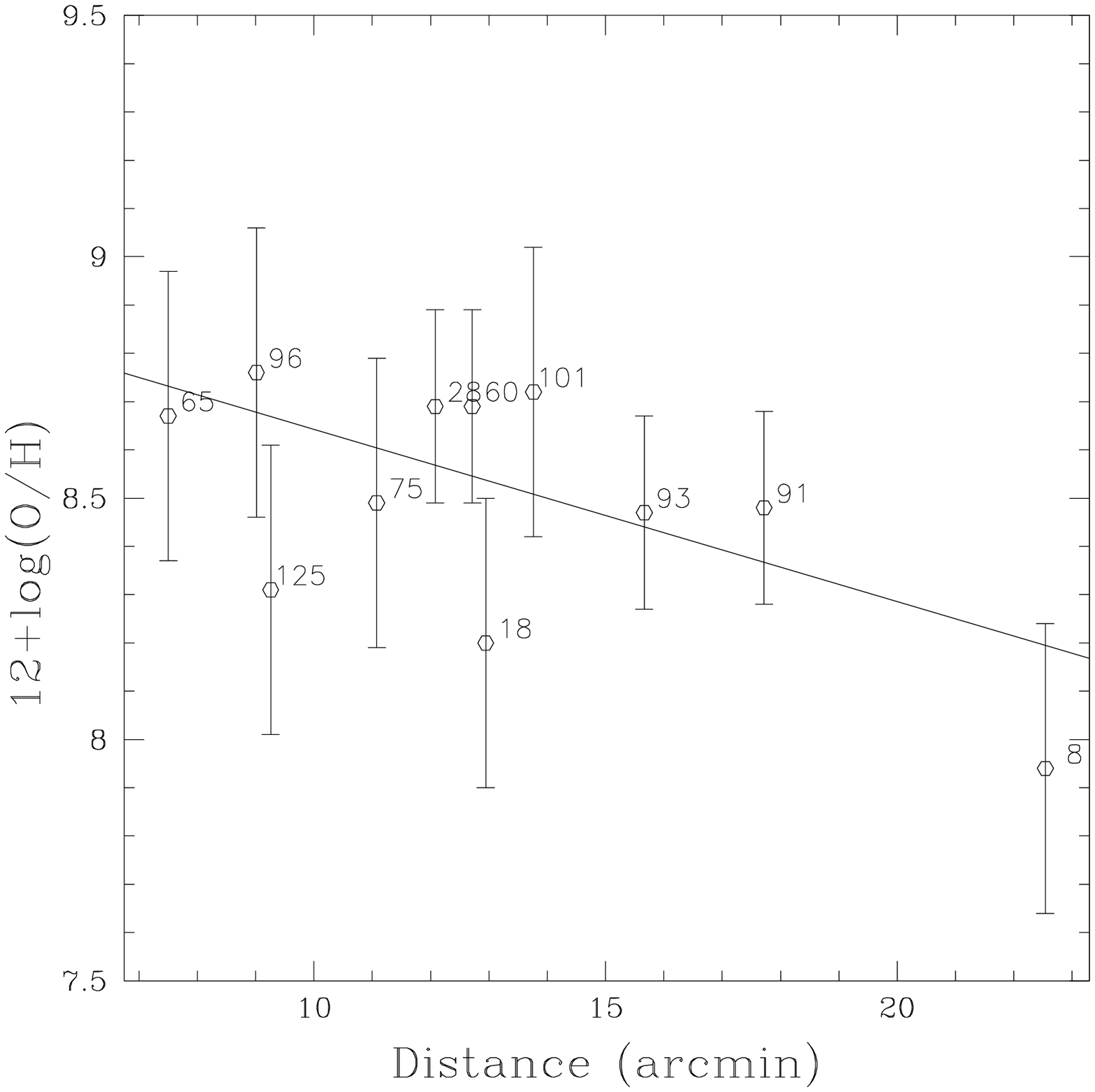}} 
\caption{ The galactocentric trend of oxygen in M~33 from 
11 PNe, recognized by their identification numbers in Table \ref{tab1}. 
The solid line is the galactocentric gradient found from \hii\ 
regions by V\'{\i}lchez et al. (\cite{vilchez88}) in the same galaxy.  }
\label{fig3b}
\end{figure}

In Fig.\ref{fig3b} we present the derived oxygen abundances  as function of the galactocentric 
projected distance. We find a possible consistency with oxygen gradient across the disc 
of M~33 detected by V\'{\i}lchez et al. (\cite{vilchez88}) 
for \hii\ regions. 
A weighted least squares fit gives a 
$\Delta \log$(O/H)/$\Delta$d=-0.14~dex kpc$^{-1}$, assuming the distance of 840~kpc, 
to be compared with the overall gradient by V\'{\i}lchez et al. (\cite{vilchez88}), $\Delta \log$(O/H)/$\Delta$d=-0.10~dex kpc$^{-1}$, 
considering the same distance. The Pearson correlation coefficient for our relationship is 0.6.
ON the other hand our result is quite dependent on a single PN (n. 8). Therefore 
the study of more PNe at high galactocentric distances is needed to clarify this aspect.

\begin{table}
\caption{Comparison among CLOUDY  both with blackbody central star and a constant density, with model atmosphere by Rauch (\cite{rauch}), and with BB and a density dependent on r$^{-2}$ (two PNe)  abundances determinations. The last raw gives the ICFs abundances determinations (M03).}  
\label{tab2}
\begin{center}
\begin{tabular}{r l r r r r r}    
\hline   
Id & Method & He/H & N/H & O/H & S/H & Ar/H \\
\hline
91 & CL$_{BB}$	        & 11.31 &7.86 &8.61 &6.42 	 &6.26\\
   & CL$_{rauch}$	& 11.20 &8.01 &8.64 &6.52	 &6.25\\
   & CL$_{BB+dens-r^{-2}}$ & 11.31 &8.07 &8.48 &6.60        &6.18\\ 
   &ICF$_{\rm {M03}}$	& 11.28 &7.83 &8.70 &6.61$\dag$  &6.32\\
\hline 
93 & CL$_{BB}$               & 11.00 &7.53 &8.69 &6.46	      &6.17\\
   & CL$_{rauch}$     	     & 10.92 &7.55 &8.60 &6.44	      &6.13\\	
   & CL$_{BB+dens-r^{-2}}$             & 10.93 &7.75 &8.47 &6.66        &6.09\\
   &ICF$_{\rm {M03}}$       & 11.00 &7.50 &8.85 &6.44$\dag$ &6.36\\
\hline
101 & CL$_{BB}$               & 11.07&7.33 &8.76 &6.18       &6.27\\
    & CL$_{rauch}$     	      & 11.14&7.34 &8.70 &6.13       &6.22\\
    &ICF$_{\rm {M03}}$        & 11.15&7.21 &8.72 &6.32$\dag$ &6.45\\
\hline
\end{tabular} 
\end{center}
$\dag$See text.
\end{table}

\subsection{Relationship among chemical abundances}
\begin{figure*}
\resizebox{\hsize}{!}{\includegraphics{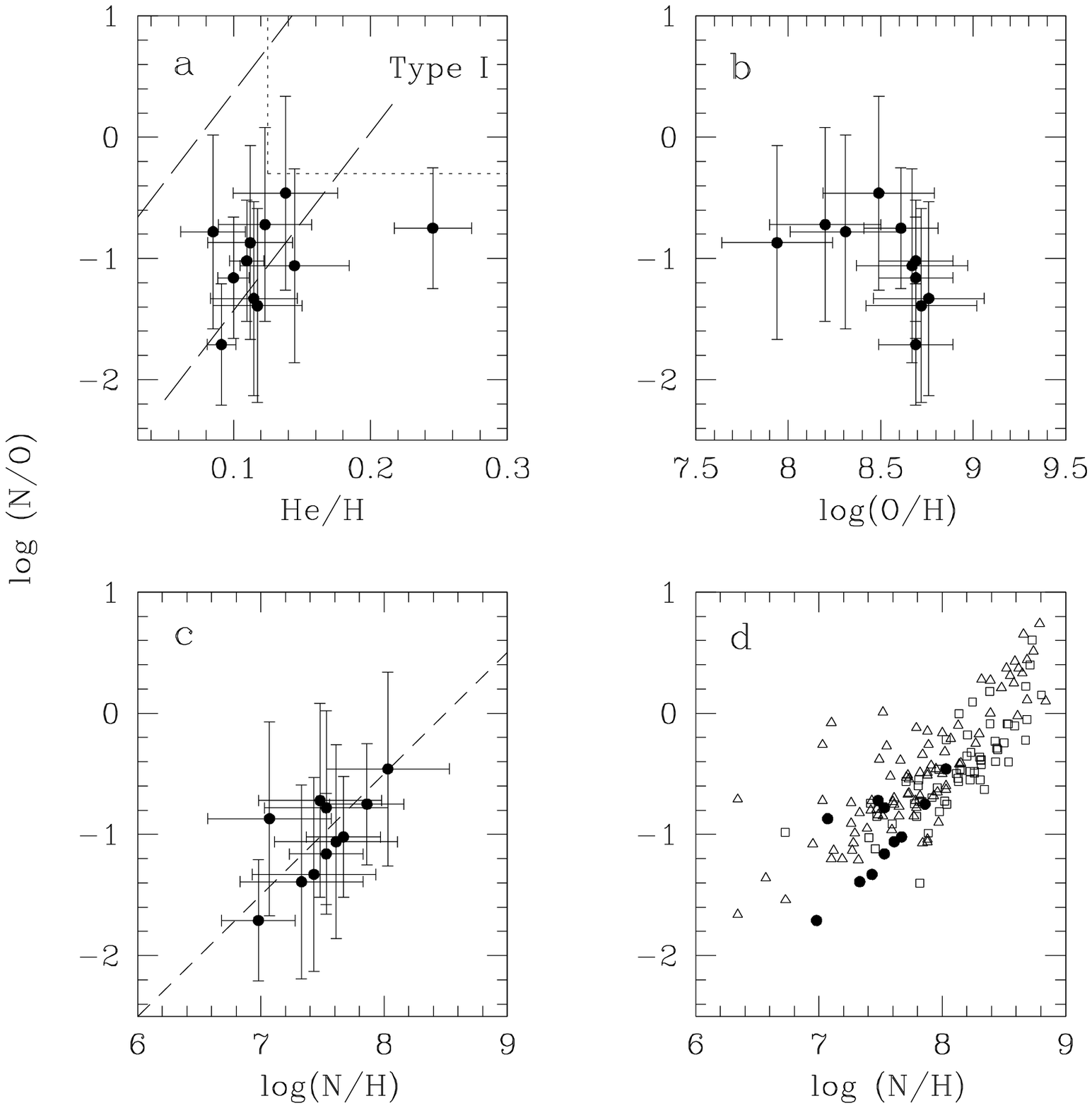}} 
\caption{Some relationships between chemical abundances. a) He/H {\sl vs.} N/O.
The area included inside the dashed lines is the main locus of
Galactic PNe derived from Perinotto et al \cite{p04}, hereafter P04,
Fig.3. The dotted lines show the limits of type I PNe area. b) O/H {\sl vs.} N/O: 
an anti-correlation in shown as seen by
JC99 in M~31; M~33 PNe (present paper) appear to reach lower N/O values than
M~31 PNe. c) N/H {\sl vs.} N/O. The dashed line shows the Galactic
relation derived by Henry (\cite{henry}) as shown by JC99.  In spite
of some dispersion, the Galactic relation between N/H and N/O appears
to apply as well to M~33. d) N/H {\sl vs.} N/O. Filled circles are PNe
of M~33 (this work), triangles are PNe of LMC (Leisy et al. \cite{leisy04}) 
and squares are Galactic PNe (P04). }
\label{fig3}
\end{figure*}
 
In Fig.~\ref{fig3}a we show the relation between He/H and N/O.  
Chemical abundances are plotted with their errors.  The area included
inside the dashed lines is the locus of Galactic PNe,  derived from
Fig.3 in P04.  
 Following the recent results  by  Exter et al. \cite{exter} 
for both bulge and disc Galactic PNe on  N/O vs He/H,  PNe with  
low N/O show also low He/H, but at N/O $>$ 0.25 the whole range of He/H 
is sampled. 
The existence of a relationship between He/H and N/O  has been discussed thoroughly by KB94, 
who found no correlation, although a correlation in the Type I PNe 
alone might exist. This issue is  difficult to assess
because there are intrinsic problems  in determining helium abundance such as 
the collisional corrections and the uncertain ionization fractions. 
The data of M~33 for N/O vs He/H suggest a behaviour similar to the galactic one, but 
further studies are in order to really ascertain that.     

Fig.~\ref{fig3}b shows O/H {\it vs} N/O. 
 The existence of an anticorrelation between these quantities is particularly controversial 
in the literature. Our data indicate a weak anti-correlation, with a  Pearson correlation coefficient 
of 0.5 (0.2 in case of a weighted anti-correlation). 
Moreover, it appears that, for
12+$\log$(O/H)$<$8.5, only high values of N/O are found, while there is
a larger spread for higher O/H abundances. 
The existence of this effect  has also 
been found in Galactic bipolar PNe (Perinotto \& Corradi
\cite{per98}), in the Galactic bulge (Stasinska et al. \cite{stasinska}), 
in the SMC (Costa et al. \cite{costa}, Leisy \& Dennefeld \cite{leisy04}), possibly in M31
(JC99) and the LMC (Leisy \& Dennefeld \cite{leisy04}), although it is not
apparent neither in the sample of Galactic PNe by KB94 (covering the range
8.2$<$12+$\log$(O/H)$<$9.2) nor in bulge and  disc Galactic PNe 
by Exter et al. \cite{exter} (8$<$12+$\log$(O/H)$<$9.2).

In the cases where this transition to
high N/O ratios for low oxygen abundances is well visible, it occurs
at different O/H for different galaxies, and roughly at the average
overall metallicity computed from other classes of objects like HII
regions.  This suggests that the explanation offered by Costa et
al. \cite{costa}, i.e. the dredge-up increasing N/O
in the AGB envelope is more efficient at low metallicities, might be not 
the main one, as the transition to high N/O ratios would then
occur always at the same metallicity. Together with the fact that in
the Galaxy the effect is better visible for high mass progenitors like
bipolar PNe (Perinotto \& Corradi \cite{per98}), the data would instead point
to a high efficiency of the ON cycle in the most massive PN
progenitors (Henry \cite{henry}), that would lower the O/H abundances
while increasing N/O.

In Figs. \ref{fig3}c and \ref{fig3}d the relationship  between N/O and
N/H are presented. In Fig. \ref{fig3}c the dashed line represents the
correlation found by Henry (\cite{henry}) for the Galactic PNe.  Our
data, shown in Fig. \ref{fig3}c with their errors, show some scattering, 
but the Galactic relation between N/H and N/O appears to
apply as well to M~33 PNe. The Pearson correlation coefficient is 0.7. 
JC99 suggested the meaning of this
correlation: its slope is close to unity and thus the variations
of N/O are mainly due to the variation of N/H (due to nucleosynthesis
of N from progenitors of different mass), while oxygen abundance changes 
less during the life of PN central stars.  In
Fig.~\ref{fig3}d, N/O {\sl vs.} N/H is reported for Galactic PNe
(P04), for PNe of the LMC (Leisy \& Dennefeld \cite{leisy04}) and of M~33
(this paper).  A similar trend is common to PNe of these three
galaxies.  

 We come now to the number of Type I PNe in our sample. 
The Type I PNe, according to the definition of Peimbert \& Torres-Peimbert
(\cite{peimbert}) are those with $\log$(N/O)$\ge$-0.3 and high He/H$\ge$0.125. 
A more stringent definition has been proposed by KB94,  $\log$(N/O)$>$-0.1.
On the other hand it has to be noted
that the definition of Type I PNe should be a function of metallicity, because the amount of N
that can be produced by hot bottom burning is dependent on the amount of C present 
and  also because the O/H abundance depends on the metallicity of the galaxy.
Thus a different criterion should be adopted for Type I PNe in a galaxy of different 
metallicity from our Galaxy. The N/O limit above which a PN is a Type I, meaning 
that it has experienced hot bottom burning conversion of C to N, should be the C$+$N/O 
ratio from the \hii\ regions. This is difficult to estimate since values for C are hard to obtain. 
Thus, since the metallicity of M~33 is not so different to that of our Galaxy, we still 
apply the Galactic criterion to discriminate Type I PNe. 
From Fig. \ref{fig3}a, we note a lack in M33 of
type I PNe, using the  Peimbert \& Torres-Peimbert
(\cite{peimbert}) criterion. Moreover, we find no Type I PNe with the definition by KB94.

The average $\log$(N/O) of M~33 PNe is -0.89, see Fig. \ref{fig3}a.  
This value is  lower than the Galactic non-Type I
PNe (KB94) value (-0.55) and it is close to the non-Type I PNe
value (-0.90; Leisy \& Dennefeld \cite{leisy04}) of the LMC, which is a galaxy of similar metallicity 
(see van den Bergh \cite{bergh}), whereas Galactic and
LMC Type I PNe have a higher value (+0.07 and +0.02, respectively, see
Table \ref{tab1}).

 The number of Type I PNe might be connected to the metallicity of the host galaxy. 
In a deep survey for PNe of the SMC by Jacoby \& De Marco (\cite{jacobydemarco}), 
they found an extremely large number of PNe where the \nii\ lines are stronger 
than \ha. This might translate to a large Type I PN fraction. 
Comparing the fraction of Type I PNe also in the Galaxy and in the LMC, they
determined that there might be a trend of more Type I PNe for lower 
metallicity galaxy. They explained this trend with the fact that more 
massive PN central stars, which are progenitors of Type I PNe, 
produce more dust and thus end up being systematically dimmer and 
consequently harder to detect. 
This would be more so on high metallicity galaxy since more dust should 
be produced. 
M~33, which has a metallicity intermediate between the Galaxy and the LMC, 
might be expected to have a similar number of Type I PNe (between $\sim$17\% and $\sim$20\% 
cf.  Jacoby \& De Marco (\cite{jacobydemarco})).

An explanation for the smaller number of Type I PNe detected might be that Type I PNe, 
which are principally represented by bipolar morphological
type and generally associated with younger and more massive
progenitors (cf. Corradi \& Schwarz \cite{corradi}), are less
luminous in the \oiii\ 5007 \AA\ line than other morphological types
of PNe (e.g. Magrini et al. \cite{m04}, see Figs. 1 \& 2, from which one notes that
the cutoff of bipolar PNLF is fainter $\sim$1.5 mag for LMC and
$\sim$2 mag for the Galaxy than that of elliptical PNe). 
We remind the reader that the PNe presented in this paper
have been discovered with a survey complete up to $\sim$2~mag 
fainter than the bright edge of the luminosity 
function (Magrini et al. \cite{m00}).  
If we consider that the results of luminosity functions of PNe with
different morphology presented by Magrini et al. \cite{m04}, especially
in the case of the Galaxy which has a similar morphology (M~33 is Sc III and the Galaxy is
S(B)bc I-II, cf. van den Bergh \cite{bergh}), apply to M~33 as well,
this means that  a number of bipolar Type I PNe could have 
been lost because of the incompleteness of the survey; i.e. one should expose more in \oiii\ 
to detect Type I PNe.

\subsection{Oxygen abundance and  {\rm [O~III]} luminosity}

\begin{figure}
\resizebox{\hsize}{!}{\includegraphics{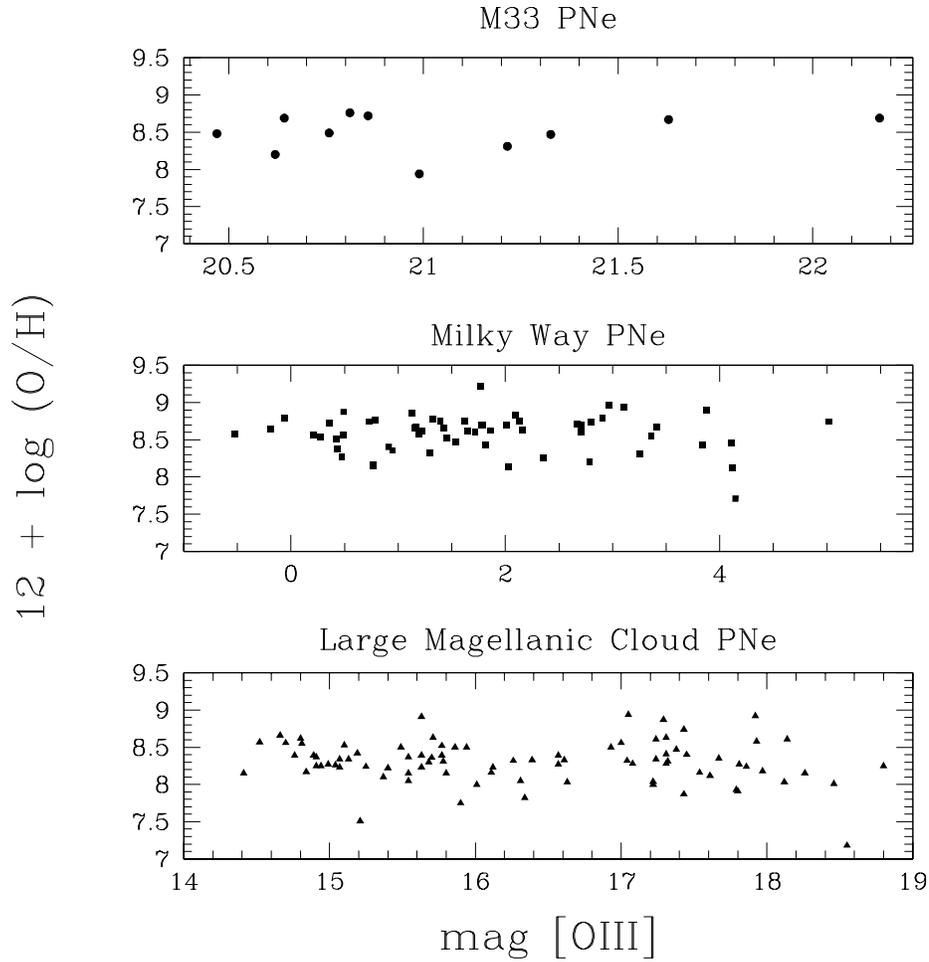}} 
\caption  {\oiii\  luminosity {\sl vs} O/H for M~33
PNe (upper panel), MW PNe (central panel) and LMC PNe (lower panel).}
\label{fig4}
\end{figure}

The determination of chemical abundances in an external galaxy rests
on its brightest objects.  We have examined the O/H {\sl vs}
\oiii\ luminosity in the M~33 sample, in the LMC, and in the Galaxy to test for
possible biases in measuring O/H from the brightest PNe (see
Fig.\ref{fig4}).  The sample of Milky Way PNe consists of objects in
common between PNe whose chemical abundances have been re-determined
by P04 and PNe whose distances have been recently obtained by Phillips
(\cite{phillips}), whereas the LMC sample consists of PNe with chemical
abundances and relative fluxes from the work of Leisy \& Dennefeld
(\cite{leisy04}), and the absolute fluxes from Jacoby et
al. (\cite{jacoby90}). 

One could expect that there would be a
correlation between the oxygen abundance and the \oiii\ 5007 \AA\
luminosity.  However no trend of O/H {\sl vs} \oiii\ luminosity is seen 
in the three galaxies, as already noted for M~31 by JC99.  
There might be various reasons for that.  The basic
point is that oxygen is one of the most important coolants of
the nebula so that a high O concentration lowers the nebular
temperature and consequently the flux coming from the \oiii\ 5007 \AA\
line.  There is therefore a feedback effect between a higher O/H
concentration and a consequently lower T$_{e}$ which produce the
mentioned lack of trend of O/H {\sl vs.} \oiii\ luminosity.  It means
that O/H derived from the brightest PNe (as from any PN) is
representative of the total PNe population. This is particularly
useful when deriving chemical abundances from PNe belonging to
external galaxies, where only the brighter tip of their luminosity
function is observed.

\section{Summary and Conclusions}

Using the CLOUDY photoionization code, we have derived nebular parameters 
and some stellar effective temperatures for  11 PNe belonging to the spiral galaxy M~33, and
measured their He/H, O/H, N/H, Ar/H and S/H abundances.
 The data are consistent with 
the  O/H gradient from \hii\ regions by V\'{\i}lchez et al. (\cite{vilchez88}), 
but abundances in PNe farther from the centre of M~33 than the sample 
studied here are needed  to ascertain a clear  metallicity gradient from PNe in M~33.
A trend of N/O to N/H similar to the Galactic ones has been noted.  
We did not find PNe with the very high N/H
and N/O values that are typical of galactic Type I PNe.  
A possible explanation is that the
incompleteness of the survey for M33 PNe (M00) caused the loss 
 of  Type I PNe (N/O$>$-0.3), which are generally faint in the
\oiii\ 5007 \AA\ emission line. 
Finally we found that the \oiii\ luminosity is clearly independent of the oxygen abundance.  Therefore
it is possible to use every PN, and particularly the brightest PNe, in
external galaxies as representative of the whole PN population.

\begin{acknowledgements}    
We especially thank Dr. Roberto Baglioni for his precious help 
in writing the Perl procedure. We are extremely grateful to an anonymous referee 
for comments and  useful suggestions that improved significantly the  paper.
\end{acknowledgements}


\begin{thebibliography}{}
\bibitem[2002]{armsdorfer}
Armsdorfer B., Kimeswenger S., \& Rauch T., 2002, 
RMxAC, 12, 180 I
\bibitem[1932]{ambar}
Ambartsumian V. A., 1932,
Poulkovo Obs. Circ 4, 8
\bibitem[1983]{che}
Che A. \& K$\ddot{o}$ppen J. 1983, 
A\&A, 118, 107
\bibitem[1995]{corradi}
Corradi, R. L. M. \& Schwarz, H., 1995,
A\&A, 293, 871
\bibitem[2000]{costa}
Costa, R. D. D., de Freitas Pachego, J. A., \& Idiart T. P., 2000,
A\&AS 145, 467
\bibitem[1989]{dennefeld}
Dennefeld, M., 1989,
in {\sl Recent Developments of Magellanic Cloud Research}, eds.  K.S. de Boer, F. Spite, G. Stasinska,
 Publ. Observatoire de Paris,  Meudon, 107
\bibitem[1984]{dufour84}
Dufour, R. J. 1984, In {\sl IAU Symposium 108, Structure and evolution
of the Magellanic Clouds} ed D. Reidel (Dordrecht:Reidel), 353
\bibitem[2004]{exter}
Exter K. M., Barlow M. J., \& Walton N. A., 2004, 
MNRAS, 349, 1291
\bibitem[1998]{ferland}
Ferland G. J., Korista K. T., Verner D. A.,
 Ferguson J. W., Kingdon J. B., Verner E. M., 1998,
PASP, 110, 761
\bibitem[1988]{gurzadyan}
Gurzadyan G. A., 1988,
ApSS, 149, 343
\bibitem[1996]{gur97}
Gurzadyan G. A., 1997,
``The Physics and Dynamics of Planetary Nebulae''  (Berlin: Springer), p. 112
\bibitem[1990]{henry}
Henry R. B. C., 1990,
ApJ, 356, 229
\bibitem[1997]{howard}
Howard J. W., Henry R. B. C., \& McCartney S., 1997, 
MNRAS, 284, 465
\bibitem[1983]{iben}
Iben I. Jr. \&  Renzini A., 1983,
ARA\&A, 21, 271
\bibitem[1999]{jc99}
Jacoby G. H. \&  Ciardullo R., 1999,  
ApJ, 515, 169 (JC99) 
\bibitem[2002]{jacobydemarco}
Jacoby G. H. \& De Marco O., 2002,
AJ, 123, 269
\bibitem[1990]{jacoby90}
Jacoby, G. H., Walker, A. R. \& Ciardullo, R., 1990,
ApJ, 365, 471
\bibitem[1994]{kb94}
Kingsburgh R. L. \& Barlow M. J., 1994,
MNRAS 271, 257
\bibitem[2004]{leisy04}
Leisy P.\&  Dennefeld, M., 2004,
in preparation
\bibitem[2000]{leisy}
Leisy P., Dennefeld M. \& Francois, P., 2000,
in Proceeding of Ionized Gaseous Nebulae. Mexico City November 21-24, 2000, p. 32
\bibitem[2000]{m00}
Magrini L., Corradi R. L. M. \& Mampaso A., Perinotto M.,  2000,
A\&A, 355, 713 (M00)
\bibitem[2003]{m03}
Magrini L., Perinotto M., Corradi R. L. M. \& Mampaso A., 2003,
A\&A, 400, 511 (M03)
\bibitem[2004]{m04}
Magrini L., Corradi R. L. M., Leisy, P. et al., 2004,
in ``Asymmetric Planetary Nebulae III'', ASP conf. series, in press
\bibitem[1984]{maran}
Maran S. P., Gull T. R., Stecher T. P., Aller L. VH. \&  Keyes C. D., 1984,
ApJ, 280, 615
\bibitem[2004]{morisset}
Morisset, C., 2004, 
AJ, 601, 858
\bibitem[2003]{rauch}
Rauch, T., 2003,
A\&A 403, 709
\bibitem[1995]{richer}
Richer M.G. \& McCall M., 1995,  
ApJ, 445, 642
\bibitem[1983]{peimbert}
Peimbert M. \& Torres-Peimbert S., 1983,
in IAU Symp. 103, Planetary Nebulae, ed. D.R. Flower (Dordrecht:Reindel), p. 233
\bibitem[1998]{per98}
Perinotto M. \& Corradi R. L. M., 1998, 
A\&A, 332, 721
\bibitem[2004]{p04}
Perinotto M., Morbidelli L., \& Scatarzi A., 2004,
MNRAS, in press (P04)
\bibitem[2002]{phillips}
Phillips, J. P., 2002,
ApJS, 139, 199
\bibitem[1998]{stasinska}
Stasinska G., Richer M.G. \&  Mc Call M.L., 1998, 
A\&A, 336, 667
\bibitem[2000]{bergh}   
van den Bergh S. 2000 in {\em The Galaxies of the Local   
Group}, Cambridge University Press   
\bibitem[1988]{vilchez88}
V\'{\i}lchez J.M., Pagel B.E.J., Diaz A.I., Terlevich E., Edmunds M.G. 1988, 
MNRAS 235, 633
\bibitem[1997]{walsh}
Walsh J. R., Dudziak G., Minniti D., \& Zijlstra A. A., 1997,  
ApJ, 487, 651  
\bibitem[1999]{walsh99}
Walsh J. R., Walton N. A., Jacoby G. H. \& Peletier R. F., 1999,  
A\&A, 346, 753  
\end{thebibliography}
\end{document}